# Evolution of Transthyretin

J. C. Phillips

Dept. of Physics and Astronomy, Rutgers University, Piscataway, N. J., 08854

Abstract

Evolution of the amino acid sequences of transthyretin (TTR) can provide additional information about its dynamics that both complements and extends the already extensive static structural data. Protein dynamics is largely driven by interactions between the protein itself and the thin water film that covers it. Here those interactions are connected to profiles of water waves connecting domain pivots (hydrophobic extrema), as well as a broad hydrophilic central hinge region that has evolved to provide human TTR with greater flexibility and stability. This central region has a density of single mutations related to amyloid polyneuropathy that is three times higher than other regions.

Introduction

Transthyretin (TTR) is a transport protein in the serum and cerebrospinal fluid that transports the thyroid hormone thyroxine ($T_4$) and retinol to the liver. The liver in turn secretes TTR into the blood, and the choroid plexus secretes TTR into the cerebrospinal fluid. TTR misfolding and aggregation is known to be associated with amyloid diseases. The molecular structure is a homotetramer with a dimer of dimers. Each monomer contains 127 amino acids, and is stiffened by a rich beta sheet structure. The overall structure resembles hemoglobin in both size and shape, but the hemes are missing. Human TTR is described in detail at Uniprot P02776, where 75 single mutations are listed, nearly all related to amyloid polyneuropathy.

Here we apply a new method of implementing molecular dynamics that has successfully analyzed the evolution of function of many proteins, including hemoglobin [1] and more recently, Coronavirus spikes (~ 1200 amino acids), where evolution from CoV-1 to highly contagious CoV-2 involves ~ 300 mutations [2], as well as Omicron (30 more mutations) [3]. The new method is post-Newtonian (it does not involve force fields); it implements instead the concept of evolutionary approach to a phase-transition critical point [4], whose properties are described in terms of fractals [5] fitted to curvatures of > 5000 protein segments [6]. Just as Nature has designed life around only 20 amino acids, so has water built the topological shapes of protein structures around these 20 fractals, denoted by Ψ(aa). While introductions to these post-Newtonian methods are given in all papers, physicists may find an early one on Hen Egg White most accessible [7]. Broadly speaking, the new method utilizes keys that break the amino acid code [8].

Results

The present post-Newtonian molecular dynamics relies primarily on amino acid sequences, with at most only passing references to structures. The advantage of this simple approach is that it can describe evolutionary sequences, which have become available in the 21$^{st}$ century [9]. There are many structural studies of human TTR, and few for rat and chicken. The first step in analyzing a protein sequence is choosing the best sliding window of width W over which the fractal hydropathicities are averaged. One can guess that evolution will develop in such a way as to enable protein domains to collaborate in functioning more rapidly. These domains would be centered on hydrophobic extrema, and separated at their edges by hydrophilic extrema of Ψ(W,aa). Such domains are dynamical domains; while they often resemble static domains, their structure may be more detailed. Is there a way to average Ψ(aa) over a suitably chosen length W? Of course, that is exactly what a sliding window does; the result is a matrix, Ψ(W,aa). Sliding windows are often used to smooth long sequences of data, without knowing many details, but here we already know quite a lot about static domains. This situation gives the theorist a chance to imagine himself as a master builder. If he were engineering evolution of a given protein through small differences that leave its fold unchanged, what would he do?

It seems likely that proteins have evolved to be either more or less compact. A simple way to measure in/out compactness is the range of Ψ(W,aa) over the entire protein is through its variance V (the mean of its square – the square of its mean). Now by plotting V(A,W, Ψ(aa))/ V(B,W, Ψ(aa)) as a function of W, we can find features (such as extrema) consistent with functional evolution. The variance ratios for human and chicken are shown in Fig. 1 for two Ψ(aa) scales, the MZ [5] and KD [10] scales. The MZ scale is associated with second-order phase transitions, whereas the KD scale is associated with unfolding (first-order transitions). The differences studied here are all second-order, and the MZ curve in Fig. 1 has a clear peak at W = 13. This confirms the critical and continuous (second order)l nature of TTR evolution. It is also encouraging that the phase transition method does not involve first order unfolding, which is included in leading methods of Newtonian simulations [11].

Given W = 13, we compare human and chicken Ψ(13,aa) in Fig. 2. The most obvious differences between human and chicken are the deeper hydrophilic edges in human near 22, 65, and 119. The deepest hydrophilic edge near 65 is near the monomer center, which makes the protein more flexible overall in terms of an N part 21-60 and a C part 70 – 130 "dimer". Outside edge flexibilty facilitates tetramerization.

In studies of other proteins we often found multiple level edges, including up to 12 in Omicron [3]. Level sets have been used by James Sethian, an applied mathematician, to study interface propagation and synchronized assembly in many systems, including fluid mechanics, semiconductor manufacturing, industrial inkjets and jetting devices, shape recovery in medicine, and medical and biomedical imaging [11]. He also uses the ancient Voronoi geometrical construction to separate interfaces [12], much as [5] did to separate amino acids. Overall, level edges are a powerful tool that enables us to identify domain synchronization. Here one readily identifies the four nearly level hydrophobic edges, with (1,2) in the N part, and (4,5) in the C part

Uniprot also contains TTR for rat. As shown in Fig. 3, rat TTR is close to human, with no deletion. The differences are interesting and systematic. . Although edge 3 is only weakly hydrophobic, its hydrophobicity has increased substantially in human compared to rat, providing additional stability to human TTR. There the human-rat effect is larger than it was for human-chicken; in the latter case, chick is much less hydrophilic at the N terminal.

The large number (at least 75) of mutations related to amyloid polyneuropathy at first discourages analysis, but simply counting the number of mutations in each site decade shows that the region 50-80 contains at least 12 mutations/decade, while outside this region the number drops to 4±1. This region matches the broad hydrophilic region in Fig. 2. Its depth, relative to hydroneutral at 155, nearly doubled from Chick to Human, and even increased from Rat to Human (Fig. 3). These features are again suggestive of evolutionary approach to criticality. The most frequent mutation is V50M, and [14] suggested it might induce a disulfide bond by Cys30.

Conclusions

The present discussion of TTR complements the > 100 structural studies listed at Uniprot P02776. The evolution of function connects directly through level sets [13] to the evolution of the hydroprofiles shown in Figs. 2 and 3. The close connection between these profiles and the region of highest density of disease mutations is a new result, obtainable only with our new method. The overall internal "dimer" structure of a monomer is seen in TTR, but not in hemoglobin or myoglobin, as the heme modifies the shape. A curious feature of neuroglobin (the fastest globin, see Fig. 1 of [1]) is a large-scale trimeric structure, which may be a rare synchronizing factor.

Methods

Many readers are surprised to find that profiles based on interactions of amino acids with water can be highly informative and complementary to the > 100 static structures listed for natural TTR and its singly mutated variants at Uniprot P02776. The hydrodynamical methods used here are very similar to those already used on other proteins, such as [1-3,7]. Here again the most striking effects of evolution are seen in extrema of the $\Psi(13,aa)$ profiles shown in Figs. 2 and 3. Aside from the level sets in these figures, the method used here discovers a new feature in the TTR that is probably closely associated with its stable yet flexible tetrameric structure, the central hydrophilic region, which evolution has deepened. This is another factor that can inhibit TTRs dissociation, misfolding and aggregation [15]. The hydropathicity values $\Psi$ for the MZ and KD scales are listed at bioRxiv 10.1101/2021.02.16.431437.

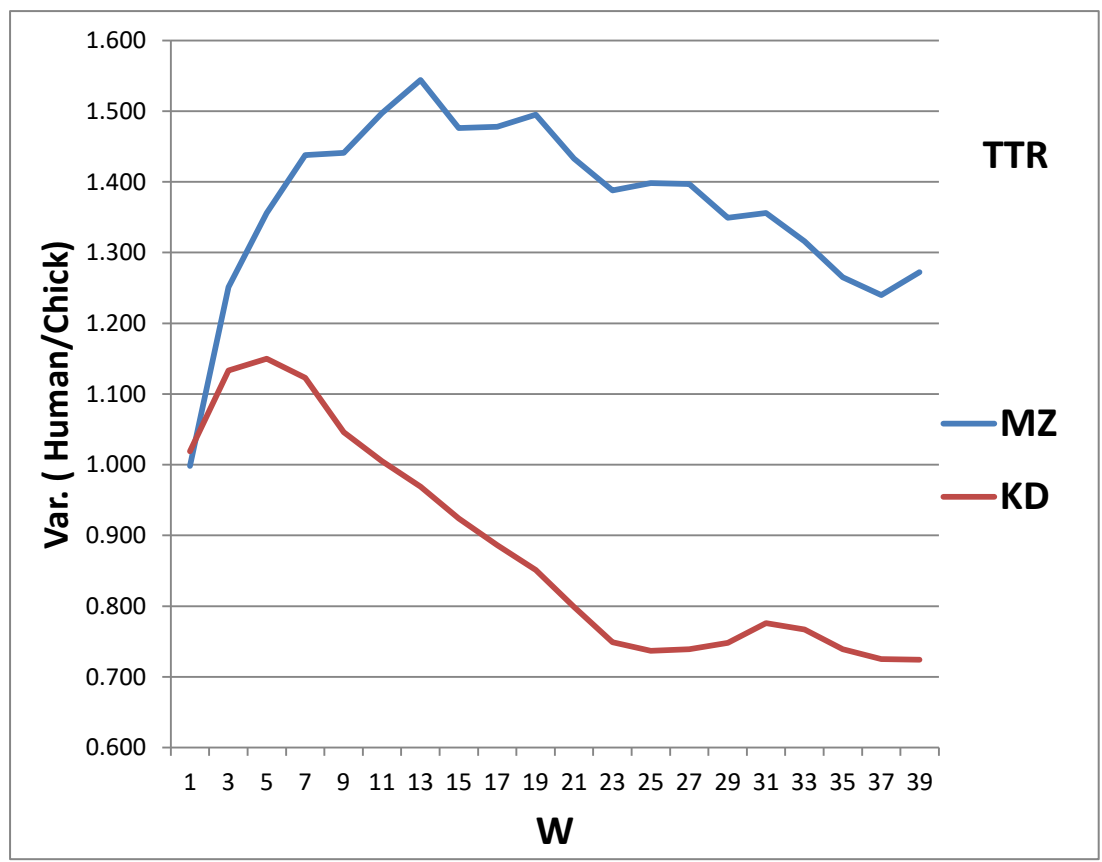

Fig. 1. The variance ratio (Human/Chick) with the first-order KD scale [8] and the second-order MZ scale [5] for TTR. The best description of the evolution of refinements in topological (in/out) shape is obtained with the MZ scale and W = 13.

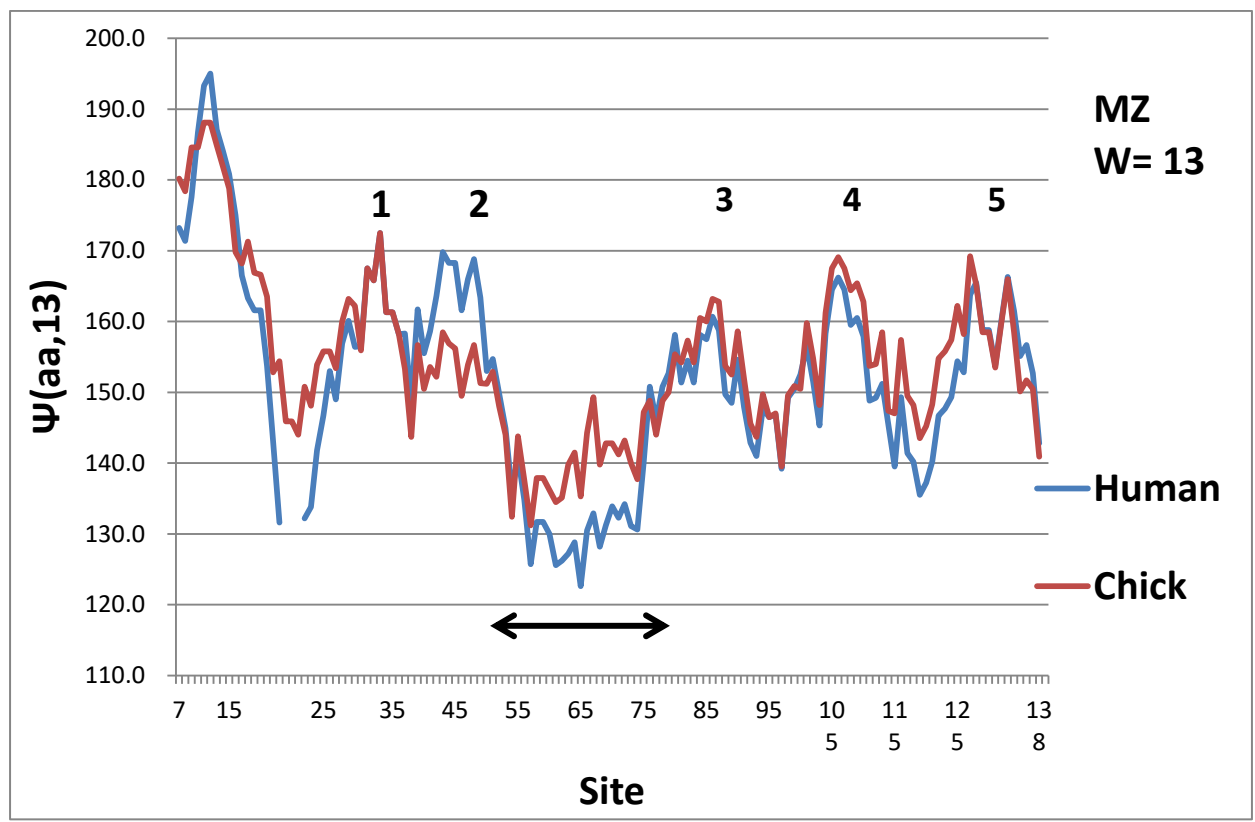

Fig. 2. Comparison of Human and Chicken hydroprofiles. Here the site numbering includes the 1-20 signal protein, which is discarded in forming the TTR tetramer. The 22LVSH25 sequence in Chicken is discarded and replaced by T22 in human, which causes the deep hydrophilic gap in the human profile. The five numbered hydrophobic edges are more level in Human than in Chicken. The double arrow marks the 50-80 hydrophilic region where there are three times more mutations responsible for amyloid polyneuropathy than in other regions (Uniprot P02766).

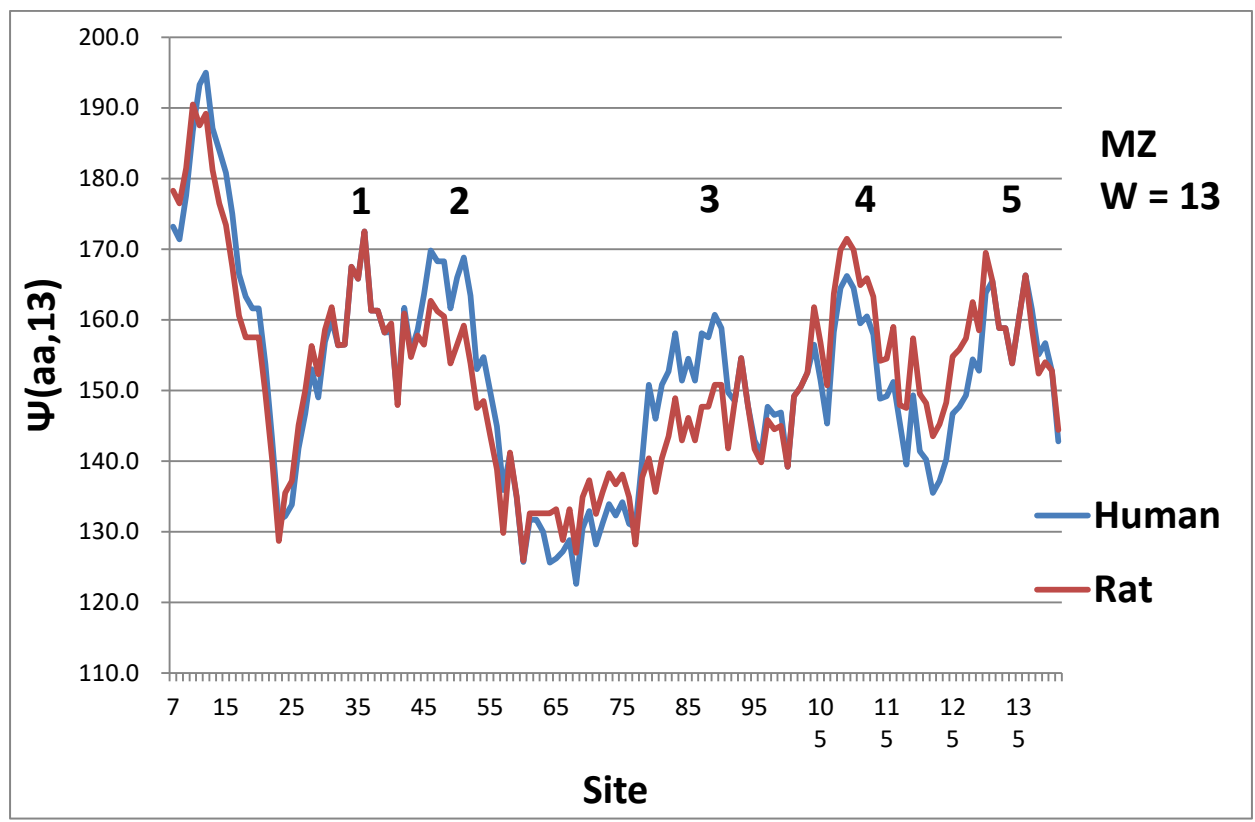

Fig. 3. Comparison of Human and Rat TTR hydroprofiles. Here again the site numbering includes the 1-20 signal protein, which is discarded in forming the TTR tetramer. There are no insertions or deletions, while the Human profile is clearly more level at the twin (1,2) and (4,5) hydrophobic edges. Also the central edge 3 has become more hydrophobic (stiffer). In the hydrophilic central region the lowest Rat edge is at 125.9, and the lowest human point is at 122.6.

|  | 36 | 45 | 87 | 105 | 127 | Ave Dev | |
|---|---|---|---|---|---|---|---|
|  | 1 | 2 | 3 | 4 | 5 | All 5 | (1,2,4,5) |
| Human | 172.5 | 169.8 | 160.7 | 166.2 | 166.3 | 3.2 | 2.5 |
| Chick | 172.5 | 158.5 | 163.2 | 169.5 | 169.2 | 4.6 | 2.7 |
| Rat | 172.5 | 162.7 | 154.6 | 171.5 | 169.5 | 6.0 | 3.2 |

Table 1. Hydrophobic edge values for three species. The surprising result that Chick is closer to Human than Rat is presumably compensation for > the deletion from Chick shown in Fig.1.